\documentclass[a4paper,amsmath,amssymb,pre,twocolumn,superscriptaddress]{revtex4-1}

\usepackage{eepic}
\usepackage{graphicx}
\usepackage{overpic}
\usepackage{latexsym}
\usepackage{bm}
\usepackage{amsmath}
\usepackage{amssymb}
\usepackage{amsthm}
\usepackage{microtype}
\usepackage[retainorgcmds]{IEEEtrantools}
\usepackage{caption}
\usepackage{subcaption}
\usepackage{array}
\usepackage{color}
\usepackage{txfonts}

\def\kcore{$k$-core}

\def\ie{\textit{i.~e.}}
\def\eg{\textit{e.g.}}

\begin{document}

\title{Analytical approach to the dynamics of facilitated spin models
  on random networks}
\author{Peter Fennell}
\affiliation{MACSI, Department of Mathematics and Statistics, University of Limerick, Ireland}
\author{James P.~Gleeson}
\affiliation{MACSI, Department of Mathematics and Statistics, University of Limerick, Ireland}
\author{Davide Cellai}
\affiliation{MACSI, Department of Mathematics and Statistics, University of Limerick, Ireland}

\begin{abstract}
Facilitated spin models were introduced some decades ago to mimic systems characterized by a glass transition.
Recent developments have shown that a class of facilitated spin models is also able to reproduce characteristic signatures 
of the structural relaxation properties of glass-forming liquids.
While the equilibrium phase diagram of these models can be calculated analytically, the dynamics are usually investigated numerically.
Here we propose a new network-based approach, called approximate
master equation (AME), to the dynamics of the Fredrickson-Andersen
model. The approach correctly predicts the critical temperature at
which the glass transition occurs. We also find excellent agreement
between the theory and the numerical
simulations for the transient regime, except in close proximity of the liquid-glass
transition. Finally, we analytically characterize the critical
clusters of the model and show that  the departures between our AME
approach and the Monte Carlo can be related to the large interface
between blocked and unblocked spins at temperatures close to the glass transition. 
\end{abstract}

\date{\today}
\maketitle

\section{Introduction}

The nature of the glass transition has been matter of debate for decades.
The key point of discussion is whether it is a purely dynamical
transition or a manifestation of a genuine thermodynamic amorphous
phase (for a review, see {\eg}
\cite{binder2011,biroli2012random,berthier2011}). In order to investigate the first hypothesis, many efforts have been spent in defining simple lattice models able to reproduce the fundamental features of the glass transition (see {\eg} \cite{berthier2011} and references wherein).
Among those, facilitated spin models (FSM), first introduced by Fredrickson and Andersen in 1984 \cite{fredrickson1984}, are perhaps the most classical simple theoretical tool able to reproduce dynamically arrested states.
It has become more and more evident, especially in experiments involving colloids, that one of the most important characteristics of glass-forming liquids is the progressive slowing of the dynamics due to the crowding of the space around each particle.
Particles spend a long time inside the cage formed by their neighbors and occasionally make a large movement to another cage \cite{weeks2000}.
A simple way to represent, albeit schematically, this caging effect in a spin model is to prescribe a geometrical constraint that hinders spin flips.
Apart from this geometrical constraint, FSMs are characterized by a trivial thermodynamics.
Despite the extreme simplicity of these models, recent developments
have shown that FSMs are even able to reproduce characteristic
signatures of the mode-coupling theory (MCT), one of the most
prominent theoretical approaches to glasses \cite{goetze2009},
including $A_2$, $A_3$ and $A_4$ singularities
  \cite{sellitto2010,sellitto2012,cellai2013a,sellitto2013}.

In spite of the relevance of FSMs, analytical study of these models has usually been focussed on the steady state, which can be calculated in simple network topologies \cite{sellitto2005}.
Regarding the dynamics, analytical approaches based on mode-coupling approximations \cite{jackle1993,pitts2001} and non-equilibrium thermodynamics \cite{garrahan2007} have been proposed, but they usually struggle in capturing the long-time relaxation of the time correlation function \cite{jackle1993,pitts2001}.
Therefore, most studies strongly rely on Monte Carlo simulations \cite{ritort2003,sellitto2005}.
However, numerical simulations become extremely slow in the proximity of the glass transition as the highly constrained kinetics has a direct impact on the speed of Monte Carlo schemes.
Therefore, an analytical approach to the dynamics of these models can
offer assistance in understanding the properties of the relaxation
process.
In this paper, we develop an accurate analytical approximation, named Approximate Master Equation (AME), of the time relaxation of the Fredrickson-Andersen (FA) model.
This approach is based on recent work \cite{gleeson2011} where
encapsulating all the nearest neighbor correlations in a master
equation provides an extremely powerful tool for a number of
binary-state models on
random networks, well beyond the mean-field approximation \cite{gleeson2013}.
We extend the master equation approach to the FA model  and show that nearest neighbor correlations are sufficient to approximate the dynamics of the model remarkably close to the glass transition.
Moreover, we identify critical clusters and show that they are
characterized by a large interface between blocked and unblocked spins.
This may explain why our approximation of the dynamics deviates from the numerical calculations in the close proximity of the glass transition.

The paper is organized as follows.
In Sec.~\ref{sec:model}, we present the FA model.
In Sec.~\ref{sec:fourmodel}, we describe the AME approach to the FA model and in Sec.~\ref{sec:results} we compare the results with Monte Carlo simulations.
Finally, in Sec.~\ref{sec:criticalclusters} we analytically characterize the critical clusters of the model and summarize our conclusions in Sec.~\ref{sec:conclusions}.

\section{The Fredrickson-Andersen model}
\label{sec:model}

The FA model~\cite{fredrickson1984} is a spin model where dynamical arrest is entirely driven by  a constraint on spin flipping based on the local neighborhood of each node.
If we  consider that each node $i$ is either in the state spin-down
$(\sigma_i=-1)$ or spin-up $(\sigma_i=+1)$, the system has Hamiltonian
\begin{equation}
{\cal H} = -\frac{1}{2}\sum_i\sigma_i.
\label{eq:hamiltonian}
\end{equation}
In addition to this thermodynamically trivial Hamiltonian, there are restrictions on spin flipping. Such restrictions
are in the form of a geometric constraint which says that spins can only flip
if at least $f$ of their neighbors are spin-down, where $f$ is called the facilitation parameter.
As a result, a spin on node $i$ flips at a rate
$W(\sigma_i \rightarrow -\sigma_i)=\min(1, e^{-\sigma_i/T})$, where $T$ is the effective
temperature of the system, if and only if the condition on the neighborhood is satisfied.
This constraint mimics \emph{caging}, a well known feature of glass-forming systems where the movements of molecules or particles, in a material close to dynamical arrest, get progressively restricted in a cage formed by the neighboring particles \cite{weeks2000}.

For further reference, we can equivalently re-write the transition rates in order to distinguish the rate $F(l_i)$ at which a node $i$
with $l_i$ spin-down neighbors changes from spin-down to spin-up from $R(l_i)$, where the opposite (from spin-up to spin-down) occurs:
\begin{eqnarray}
  F(l_i) &=&
  \begin{cases} 0 &\mbox{if } l_i < f \\
1 & \mbox{if } l_i \geq f \end{cases} \label{eq:ratefunctionF}
\\
  R(l_i) &=&
  \begin{cases} 0 &\mbox{if } l_i < f \\
e^{-1/T} & \mbox{if } l_i \geq f \end{cases}
\label{eq:ratefunctionR}
\end{eqnarray}

A relevant quantity in glassy systems is the persistence
$\phi(t)$. This is the fraction of spins that have never
flipped in the time interval $[0,t]$.
The persistence is a monotonic decreasing function of time whose long time limit
\begin{equation}
  \Phi = \lim_{t \to \infty}\phi(t)
\end{equation}
is the fraction of permanently blocked spins, and determines
whether the system is in a liquid ($\Phi = 0$) or glass ($\Phi > 0$) state.
For large temperature, $\Phi$ is zero and the system is a liquid.
As the temperature decreases, there is a critical temperature $T_c$ at which
$\Phi$ first becomes non-zero. This is the point of the
glass transition.
The FA model reproduces this transition, as well as
many features related to it, including diverging relaxation times of $\phi(t)$ close to the critical
temperature and dynamical exponents predicted by the MCT \cite{fredrickson1984,sellitto2010}.

On a degree regular tree graph (Bethe lattice), the FA model can be solved
analytically to give an expression for $\Phi$ as a function of
the system temperature $T$ for fixed facilitation $f$~\cite{sellitto2005}. 
The parameter $\Phi$
undergoes a discontinuous transition from zero (liquid) to non-zero
(glass) at the critical temperature. In this work, we build an
analytical framework that gives not only an expression
for $\Phi$, but also describes the temporal evolution of the persistence,
$\phi(t)$.

%
%


\section{The 4-state master equation approach}

\label{sec:fourmodel}

\begin{figure}[t!]
  \includegraphics[width=0.3\textwidth]{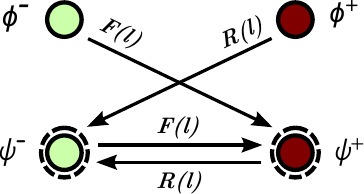}
  \caption{Schematic of the FA dynamics in the 4-state approach. The
    state of a node is a combination of spin-up or spin down and
    flipped or unflipped. Here, light green nodes are spin-down and dark red
    nodes are spin-up, while a dashed circle encompassing the
    node indicates that it has previously flipped. Nodes change from
    one state to another according to the transition rates given in
    Eqs.~(\ref{eq:ratefunctionF}) and (\ref{eq:ratefunctionR})}
  \label{fig:4stateFAtransitions}
\end{figure}

The approximate master equation (AME) formalism of \cite{gleeson2013} has been
shown to reproduce a wide range of binary-state dynamics on random networks with great
accuracy. The AME is a compartmental model where the dynamics
are described by transition rates $F_{l,m}$ and $R_{l,m}$ \footnote{The transition rates in~\cite{gleeson2013} are actually of the
  form $F_{k,m}$, $R_{k,m}$ where $k$ is the degree of a node - here we
  choose to describe them by $l=k-m$ for the sake of clarity when
  moving to the description of the 4-state dynamics} which depend
on the number ($l$ and $m$) of nearest neighbors of a node in each of
the two possible states ($-1$ and $+1$). The FA dynamics are implemented in the AME framework by taking the transition rates to be $F(l)$ and
$R(l)$ as given in Eqs.~(\ref{eq:ratefunctionF}) and
(\ref{eq:ratefunctionR}). We show in the Appendix~\ref{sec:binmodel}, however, that
considering only the spin states of each node (and using therefore a
binary AME approach) is not sufficient to capture the complexity of
the FA model. Therefore, we extend the AME approach to 4-state
dynamics by also accounting for the flipping history of each node.

\begin{table}[b!]
  \setlength{\tabcolsep}{8pt}
  \begin{tabular}{|c c c c c|}
    \hline
    State & Symbol & Spin & History & Index \\
    \hline
     $(-1,u)$  & $\phi^-$ &  $-1$  & unchanged  & $m_1$ \\
     $(+1,u)$  & $\phi^+$ &  $+1$  & unchanged  & $m_2$ \\
     $(-1,c)$  & $\psi^-$ &  $-1$  & changed  & $m_3$ \\
     $(+1,c)$  & $\psi^+$ &  $+1$  & changed  & $m_4$ \\
    \hline
  \end{tabular}
  \caption{The four possible states in the
    4-state AME approach. Index refers to the number of neighbors of a node
    in the corresponding state in the $\phi_{m_1, m_2, m_3,
      m_4}^+$ terminology discussed in the text.}
  \label{tab:states}
\end{table}

Consider a network with degree distribution $p_k$ where each node can be in one of four states
depending on its spin $(-1, +1)$ and whether or not it has
previously changed spin or are as yet unchanged $(c, u)$. These four
states are labeled $(-1, u)$, $(+1, u)$, $(-1, c)$ and $(+1, c)$ as
shown in Table~\ref{tab:states}.

\begin{figure*}[t!]
  \centering
  \begin{subfigure}[b]{0.42\textwidth}
    \includegraphics[width=0.9\textwidth]{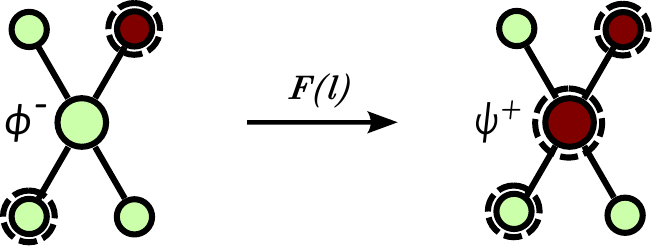}
    \caption{Node transition.}
    \label{fig:4statenodetransitions}
  \end{subfigure}
  \hspace{40pt}
  \vspace{-1pt}
  \begin{subfigure}[b]{0.42\textwidth}
    \includegraphics[width=0.9\textwidth]{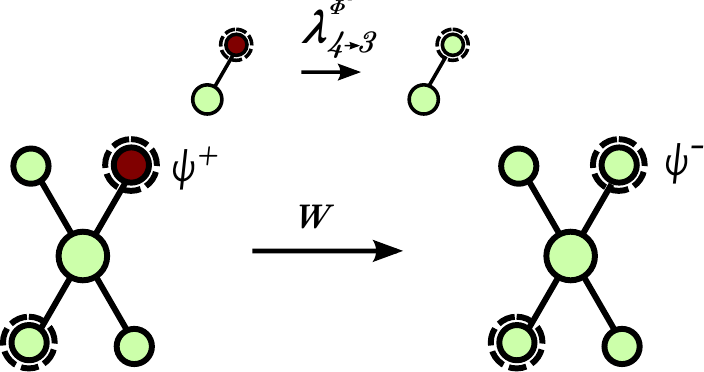}
    \caption{Neighbor transition.}
    \label{fig:4stateneighbortransitions}
  \end{subfigure}
  \caption{The 4-state AME transitions as described in the text. Node transitions are fully
    specified by the transition rates in Eqs.~(\ref{eq:ratefunctionF})
    and (\ref{eq:ratefunctionR}) while  the neighbour transition rates $W$ are
    approximated by mean field transition rates. a) This unflipped, spin-down node will
    change state to spin-up at a rate $F(l)$ where $l=m_1+m_3=3$. b) The spin-up, changed
    neighbor of the node will change to spin-down at a rate $W$ which
    is approximated by $\lambda^{\phi^-}_{4
      \rightarrow 3}$ as in
    Eq.~(\ref{eq:link_approx}).}
  \label{fig:4statetransitions}
\end{figure*}

Following the FA dynamics, nodes can change from one state to another
if the number of their neighbors $l$ that are in either of the $(-1, u)$ or $(-1,
c)$ states is at least $f$. $(-1,u)$ nodes will change
to $(+1, c)$ at a rate $F(l)$, $(+1, u)$ will change to $(-1, c)$ at a rate
$R(l)$, and $(-1, c)$ and $(+1, c)$ will change back and forth at
rates $F(l)$ and $R(l)$ respectively. This is illustrated in
Fig.~\ref{fig:4stateFAtransitions}.

Given a degree $k$ and indices $0 \leq m_i \leq k$ such that $m_1+m_2+m_3+m_4=k$, we define
$\phi_{m_1, m_2, m_3, m_4}^-(t)$ as the fraction of $k$-degree nodes in the network that are in state
$(-1, u)$ and which have
$m_1$ neighbors in state $(-1, u)$, $m_2$ neighbors in state $(+1,
u)$, $m_3$ neighbors in state $(-1,
c)$ and $m_4$ neighbors in state $(+1, c)$ at time $t$. 
The functions $\phi_{m_1, m_2, m_3, m_4}^+(t)$, $\psi_{m_1, m_2, m_3, m_4}^-(t)$ and
$\psi_{m_1, m_2, m_3, m_4}^+(t)$ are similarly defined for nodes in
states $(+1, u)$, $(-1, c)$ and $(+1, c)$, respectively. The
persistence is then given by
\begin{equation}
  \phi(t) = \Big\langle\displaystyle\sum^{\prime}_{\vec{m}	}
  	\left(\phi_{m_1, m_2, m_3,
      m_4}^-(t)+\phi_{m_1, m_2, m_3, m_4}^+(t)\right)\Big\rangle_k
  \label{eq:persistence}
\end{equation}
where $\sum^{\prime}_{\vec{m}}$ is the sum over $\vec{m}$ with the constraint $m_1+m_2+m_3+m_4=k$
and $\langle \cdot \rangle_k = \sum_{k=0}^{\infty}p_k\cdot$
symbolises averaging over the degree distribution of the network $p_k$. 

In the AME, differential equations for the system variables
are constructed by considering all flows in and out of
compartments. To illustrate, consider an unflipped node in the state
$(-1,u)$ with $m_1$, $m_2$, $m_3$ and $m_4$ neighbours in the states
$(-1,u)$, $(+1,u)$, $(-1,c)$ and $(+1,c)$ respectively. 
There is a fraction $\phi_{m_1, m_2, m_3, m_4}^-$ of such nodes in the
system. An example of a node of this type with $m_1=2$, $m_2=0$,
$m_3=1$ and $m_4=1$ is shown
in Figs.~\ref{fig:4statenodetransitions} and \ref{fig:4stateneighbortransitions}. This node will change to a different class if its state
changes from $(-1,u)$ to $(+1,c)$ (Fig.~\ref{fig:4statenodetransitions}). In
an infinitesimally small time step $dt$, this occurs with probability
$F(m_1+m_3)dt$. Thus, the fraction of nodes of this type that will leave the
compartment as a result of changing state to $(+1,c)$ in a small time step $dt$ is
\begin{equation}
  F(m_1+m_3)\phi_{m_1, m_2, m_3, m_4}^-dt
\end{equation}
Similarly, the node will leave the class if one of its neighbors
changes state (Fig.~\ref{fig:4stateneighbortransitions}). In a time step $dt$, one of its $m_4$ neighbors in the
state $(+1,c)$ will change state to $(-1,c)$ with probability
$W(\phi_{m_1,m_2,m_3,m_4}^-\rightarrow \phi_{m_1,m_2,m_3+1,m_4-1}^-)dt$. $W(\phi_{m_1,m_2,m_3,m_4}^-\rightarrow \phi_{m_1,m_2,m_3+1,m_4-1}^-)$ here is a
neighbor transition rate. Unlike the node transition rates $F$ and
$R$, the neighbor transition rates are not pre-specified. Instead,
they are 
approximated using the time-dependent link transition rates
$\lambda_{i\rightarrow j}^{\phi^-}$ as illustrated in
Fig.~\ref{fig:4stateneighbortransitions}. Thus $W(\phi_{m_1,m_2,m_3,m_4}^-\rightarrow \phi_{m_1,m_2,m_3+1,m_4-1}^-)$ is approximated by
\begin{equation}
  W(\phi_{m_1, m_2, m_3, m_4}^-\rightarrow\phi_{m_1, m_2, m_3+1, m_4-1}^-) \approx
  m_4\lambda_{4 \rightarrow 3}^{\phi^-},
  \label{eq:link_approx}
\end{equation}
where $\lambda_{4\rightarrow 3}^{\phi^-}$ is
the mean-field rate - determined by averaging over the whole network - at which links of type $(-1,u)$---$(+1,c)$ change to
$(-1,u)$---$(-1,c)$ and is given by
\begin{equation}
  \lambda_{4\rightarrow 3}^{\phi^-} = \frac{\langle\sum_{\vec{m}} m_1
    R(m_1+m_3)\psi_{m_1, m_2, m_3,
      m_4}^+\rangle_k}{\langle\sum_{\vec{m}}m_1\psi_{m_1, m_2, m_3,
      m_4}^+\rangle_k}.
  \label{eq:4statetransitionrate}
\end{equation}
The total number of nodes that will leave the
class as a result of their neighbors changing state in a time step $dt$ is
\begin{align}
m_1\lambda_{1\rightarrow 4}^{\phi^-}\phi_{m_1, m_2, m_3, m_4}^-dt+m_2&\lambda_{2\rightarrow 3}^{\phi^-}\phi_{m_1, m_2, m_3,
  m_4}^-dt \nonumber \\
+m_3\lambda_{3\rightarrow 4}^{\phi^-}\phi_{m_1, m_2, m_3,
  m_4}^-&dt +m_4\lambda_{4\rightarrow 3}^{\phi^-}\phi_{m_1, m_2, m_3,
  m_4}^-dt 
\label{eq:neighbortransitions}
\end{align}
In the other direction, nodes will enter the class as a result of
their neighbors changing state. In a time step $dt$ the number of
these will be
\begin{align}
(m_1+1)\lambda_{1 \rightarrow 4}^{\phi^-}\phi_{m_1+1, m_2, m_3,m_4-1}^-dt&  +
(m_2+1)\lambda_{2 \rightarrow 3}^{\phi^-}\phi_{m_1, m_2+1,
  m_3-1,m_4}^-dt \nonumber \\
+(m_3+1)\lambda_{3 \rightarrow 4}^{\phi^-}\phi_{m_1, m_2, m_3+1,m_4-1}^-&dt  +
(m_4+1)\lambda_{4 \rightarrow 3}^{\phi^-}\phi_{m_1, m_2,
  m_3-1,m_4+1}^-dt
\end{align}
Combining these quantities and taking the limit $dt\rightarrow 0$ results in the evolution equation
for $\phi_{m_1, m_2, m_3, m_4}^-$:
\begin{widetext}
  \begin{multline}
    \frac{d}{dt}\phi_{m_1, m_2, m_3, m_4}^- =
 -F(m_1+m_3)\phi_{m_1, m_2, m_3,
  m_4}^- \\
-m_1\lambda_{1\rightarrow 4}^{\phi^-}\phi_{m_1, m_2, m_3, m_4}^-
-m_2\lambda_{2\rightarrow 3}^{\phi^-}\phi_{m_1, m_2, m_3,
  m_4}^-
-m_3\lambda_{3\rightarrow 4}^{\phi^-}\phi_{m_1, m_2, m_3,
  m_4}^- -m_4\lambda_{4\rightarrow 3}^{\phi^-}\phi_{m_1, m_2, m_3,
  m_4}^- \\
+ (m_1+1)\lambda_{1 \rightarrow 4}^{\phi^-}\phi_{m_1+1, m_2, m_3,m_4-1}^-  +
(m_2+1)\lambda_{2 \rightarrow 3}^{\phi^-}\phi_{m_1, m_2+1, m_3-1,m_4}^- \\
+ (m_3+1)\lambda_{3 \rightarrow 4}^{\phi^-}\phi_{m_1, m_2, m_3+1,m_4-1}^-  +
(m_4+1)\lambda_{4 \rightarrow 3}^{\phi^-}\phi_{m_1, m_2,
  m_3-1,m_4+1}^-,
\label{eq:masterequation}
\end{multline}
\end{widetext}
A similar equation can be written for $\phi_{m_1,m_2,m_3,m_4}^+$. The evolution
equations for $\psi_{m_1,m_2,m_3,m_4}^-$ and
$\psi_{m_1,m_2,m_3,m_4}^+$ differ as they include nodes who enter the
class as a result of changing state - for example this extra term for the
$\psi_{m_1,m_2,m_3,m_4}^-$ variable is
\begin{equation}
  R(m_1+m_3) \phi_{m_1,m_2,m_3,m_4}^+ + R(m_1+m_3) \psi_{m_1,m_2,m_3,m_4}^+.
\end{equation} 
The full set of equations are given in Appendix \ref{sec:appendix2}. The initial conditions of this set of equations are the following. At time $t=0$, no nodes will have flipped and so 
\begin{equation}
\psi_{m_1,m_2,m_3,m_4}^-(0)=\psi_{m_1,m_2,m_3,m_4}^+(0)=0
\end{equation}
for all values $m_1,m_2,m_3,m_4$ and
\begin{equation}
\phi_{m_1,m_2,m_3,m_4}^-(0)=\phi_{m_1,m_2,m_3,m_4}^+(0)=0
\end{equation}
when $m_3>0$ or $m_4>0$. Furthermore, for the FA system with
temperature $T$ there is a fraction
$\rho=1/(1+e^{-1/T})$ of spin-up nodes at thermal equilibrium. This
gives the initial conditions on the unflipped variables for which $m_3=m_4=0$:
\begin{eqnarray}
\phi_{m_1,m_2,0,0}^-(0) &=&
p_k(1-\rho)\binom{k}{m_1}(1-\rho)^{m_1}\rho^{m_2} \\
\phi_{m_1,m_2,0,0}^+(0) &=&
p_k\rho\binom{k}{m_1}(1-\rho)^{m_1}\rho^{m_2} 
\end{eqnarray}

The master equations hold for all $m_1+m_2+m_3+m_4=k$ and for all values of $k$, resulting
in a closed system of deterministic equations from which the
expression for the evolution of the persistence is obtained:
\begin{multline}
  \frac{d\phi}{dt}=
  \Big\langle\displaystyle\sum^{\prime}_{\vec{m}} \left( \frac{d}{dt}\phi_{m_1,
      m_2, m_3, m_4}^-+\frac{d}{dt}\phi_{m_1, m_2, m_3, m_4}^+ \right) 
  \Big\rangle_k =\\
  -\Big\langle\displaystyle\sum^{\prime}_{\vec{m}} \left( F(m_1+m_3)\phi_{m_1,
      m_2, m_3,  m_4}^-+ R(m_1+m_3)\phi_{m_1, m_2, m_3, m_4}^+ \right)
  \Big\rangle_k. 
  \label{eq:persistence_evolution}
\end{multline}
The solution of the steady state is obtained by setting $d\phi/dt$
in Eq.~(\ref{eq:persistence_evolution}), and all the time derivatives of
$\phi_{m_1,m_2,m_3,m_4}^-$ and $\phi_{m_1,m_2,m_3,m_4}^+$ in
Eqs.~(\ref{eq:masterequation}) and (\ref{eq:masterequation2}), to zero.

Unlike the binary state case (see Appendix~\ref{sec:binmodel}), the 4-state AME captures the
complexities of the FA models. In the next section, we show that the value of $\Phi$
predicted by the AME corresponds to the value of $\Phi$ given
calculated by Monte Carlo (MC) simulations of the FA model. Furthermore, the evolution
of $\phi(t)$ in the MC simulations is matched well by the AME, with
the only discrepancies arising in the late relaxation of $\phi(t)$ for
temperatures close
to the glass transition.  In the final section, we explore the AME
system of equations to 
explain this discrepancy and gain an insight into the mechanism by
which the system gets stuck in the glassy state.

\section{Results}
\label{sec:results}

\subsection{Steady states}

The steady states ($t\to\infty$) of the FA model can be calculated analytically \cite{sellitto2005}.
Here we reproduce the derivation for further reference. 
For the sake of clarity, in the following calculations we consider a degree-regular graph.
However, the extension to a locally tree-like network with a generic degree distribution $p_k$ (also called `configuration model') is straightforward.
This approach is valid as in the configuration model the density of finite cycles vanishes as the network size diverges.

Let $\rho=1/(1+e^{-1/T})$ be the fraction of spin-up nodes at thermal equilibrium, as in Sec.~\ref{sec:fourmodel}.
Let us define $Z_{++}$ and $Z_{-+}$ as the probability that following an edge starting from a $\sigma = +1$ (respectively, $\sigma = -1$) spin we get to a $\phi^+$-node, {\ie} a node with $\sigma = +1$ spin which belongs to a cluster of blocked spins.
Then, the following equations hold:
\begin{equation}
	Z_{++} = \rho \sum_{l=k-f}^{k-1} {k-1 \choose l} (Z_{++})^l (1-Z_{++})^{k-1-l},
	\label{eq:Z++}
\end{equation}
\begin{equation}
	Z_{-+} = \rho \sum_{l=k-f+1}^{k-1} {k-1 \choose l} (Z_{++})^l (1-Z_{++})^{k-1-l}.
	\label{eq:Z-+}
\end{equation}
In Eq.~(\ref{eq:Z++}), the right hand side calculates all the
possibilities of having at least $(k-f)$ outgoing $\phi^+$ neighbors.
The sum in (\ref{eq:Z-+}) starts from $(k-f+1)$, instead, because on the other endpoint of the considered edge there is a $\sigma = -1$ spin, thus only $(k-f)$ outgoing $\phi^+$ neighbors would not be enough to guarantee the blockage of the considered node.
We also note that $Z_{-+}$ is just a function of $Z_{++}$.
The total fraction of blocked spins is then
\begin{equation}
	\Phi = \Phi^+ + \Phi^-,
	\label{eq:phi-from-Z}
\end{equation}
where
\begin{equation}
	\Phi^{+} = \rho \sum_{l=k-f+1}^{k} {k \choose l} (Z_{++})^l (1-Z_{++})^{k-l},
	\label{eq:phi+}
\end{equation}
\begin{equation}
	\Phi^{-} = (1-\rho) \sum_{l=k-f+1}^{k} {k \choose l} (Z_{-+})^{l} (1-Z_{-+})^{k-l}.
	\label{eq:phi-}
\end{equation}
Eq.~(\ref{eq:Z++}) has the same form of the corresponding equation for {\kcore} percolation \cite{dorogovtsev2006} and it has been shown \cite{sellitto2005,dorogovtsev2006} that the position of the phase transition can be calculated by imposing the conditions:
\begin{equation}
	\begin{cases}
		g(Z_{++})=1 \\
		g'(Z_{++})=0
	\end{cases},
	\label{eq:Z++conditions}
\end{equation}
where
\begin{equation}
	g(Z_{++}) = \rho \sum_{l=k-f}^{k-1} {k-1 \choose l} (Z_{++})^{l-1} (1-Z_{++})^{k-1-l}.
\end{equation}

\begin{figure}[t]
	\includegraphics[width=0.99\columnwidth]{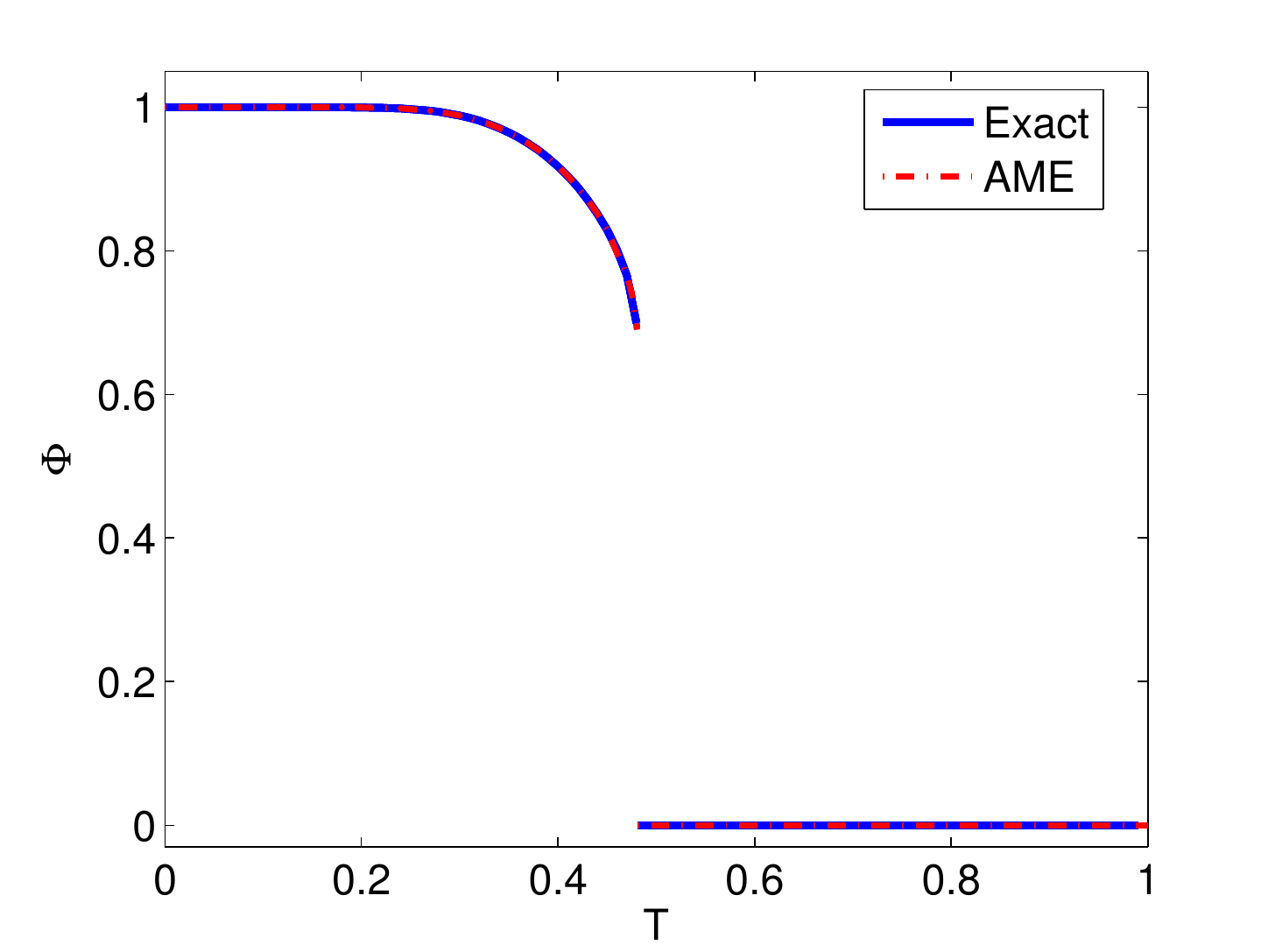}
	\caption{	Fraction of blocked spins $\Phi$ as a function of the temperature $T$.
	The blue solid line is the analytical calculation of the steady
        state as given by Eq.~(\ref{eq:Z++conditions}), while the red
        dash-dot line is calculated by our AME approach. It is evident
        that the AME predicts the exact steady state, and thus the
        critical temperature $T_c$, very accurately.}
	\label{fig:steady-state}
\end{figure}

To allow for comparison with previous results~\cite{sellitto2005,sellitto2010},  we now consider a degree regular graph with $k=4$ ($p_k = \delta_{k,4}$) and facilitation parameter $f=2$.
Solving Eq.~(\ref{eq:Z++conditions}), one can find a transition point $\rho_c=8/9$, which corresponds to the critical temperature $T_c = 1/\ln(8) = 0.480898$.
Fig.~\ref{fig:steady-state} shows the behavior of $\Phi$ at different temperatures.
At $T>T_c$, the system can relax completely after a transient regime
and there are no blocked spins in the limit $t\to\infty$.
At $T<T_c$, a finite fraction of spins remains blocked even after an infinite time.
The transition between the two phases is discontinuous with a hybrid
nature as this model is in the same universality class as bootstrap and {\kcore} percolation models \cite{chalupa1979,dorogovtsev2006,cellai2013b}.

The exact value of $\Phi$ as given by Eq.~(\ref{eq:phi-from-Z}) is compared
with the steady state values of our AME method in
Fig.~\ref{fig:steady-state}. It can be seen that the AME reproduces
the (known) steady state almost exactly, even in the proximity of the
glass transition. An implication of this is that the AME predicts the critical
temperature $T_c$ exactly. 

\subsection{Dynamics}

We now turn to the dynamics of the FA model and compare the results of
calculations from our AME approach to Monte Carlo (MC)
simulations. The MC simulations were carried out on a configuration
model network - a random network entirely described by its
degree distribution $p_k$. The network consisted of $N=2^{18}$
nodes and was updated asynchronously using a time step
of $dt=1/N$. The simulations were carried out in C/C++. The numerical integration of the AME was carried out
in MATLAB/Octave \footnote{Code for all the numerics is available
from the authors upon request.}.
As in the case of the steady state, we consider a degree regular graph with $k=4$ ($p_k = \delta_{k,4}$) and facilitation parameter $f=2$.
Because of the presence of the discontinuous transition, we expect
this case to be more challenging for our approximation with respect to
other parameter choices where the transition is instead continuous.

\begin{figure}[b]
  \centering
	\includegraphics[width=0.99\columnwidth]{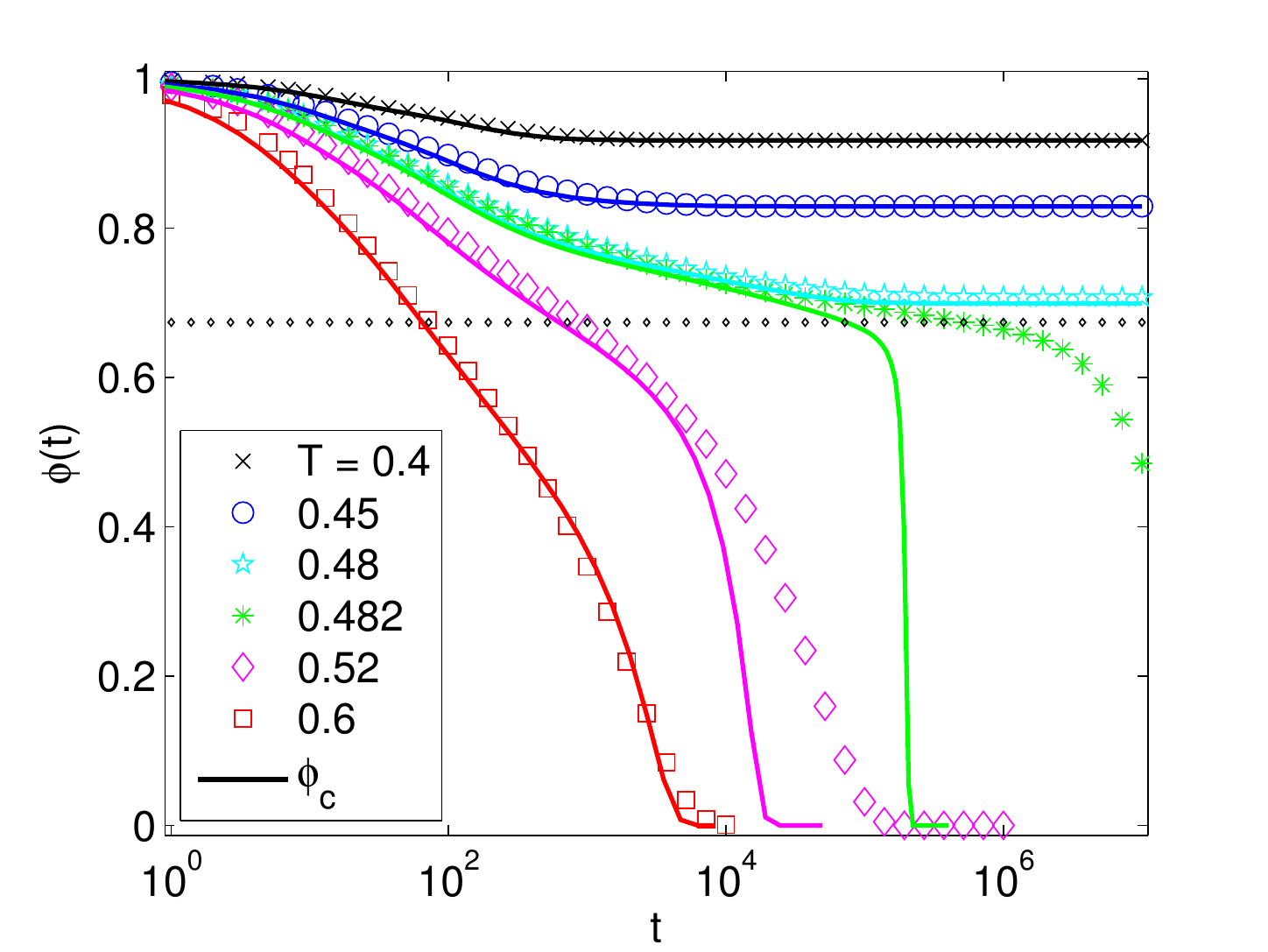}
	\caption{	Time evolution of the fraction of unflipped
          spins $\phi(t)$ for different values of the temperature
          $T$ with facilitation parameter $f=2$.
	Symbols are Monte Carlo simulations over 4-regular graphs
        (Bethe lattice) of size $N = 2^{18}$, averaged over 12 realizations.
	Continuous lines are calculated with the AME approach.
	The dotted line corresponds to the critical value of blocked spins $\Phi_c \simeq 0.69$.
	}
	\label{fig:phi-time}
\end{figure}

We consider various values of $T$ above and
below the critical temperature $T_c$. Fig.~\ref{fig:phi-time} shows the evolution of the persistence $\phi(t)$ for both the AME and
MC simulations.
Overall, we see that the AME matches the MC simulations quite well in the transient regime.
At high temperatures, the geometric constraint is less important and a detailed computation of short-ranged correlations is sufficient to capture the overall behavior of the persistence $\phi(t)$.

In the proximity of the glass transition, at $T \gtrsim T_c$, the transient regime can be
characterized by a two-step relaxation form where the two steps are
the approach and departure from the critical plateau. These are called
the $\beta$ and $\alpha$ relaxation regimes, respectively \cite{binder2011}. 
The
long-ranged correlations typical of the glass transition at this
temperature range cannot be reproduced by our AME approach, but they become
more and more important closer to the transition.
Therefore, we see the AME prediction of the $\alpha$-relaxation become
significantly less accurate as we approach the transition, despite the
fact that both the $\beta$-relaxation and steady states are correctly
reproduced as seen in Figs.~\ref{fig:phi-time} and
\ref{fig:steady-state} respectively.

At $T<T_c$, there is excellent agreement between theory and simulations
with the analytic curve reaching the exact steady state. In this
regime, it transpires, our approximation improves again as a large
portion of the network remains blocked and therefore the error in
describing the arrangements of flipping spins has a smaller
effect. It can be shown that our numerical results are robust
  against finite size analysis.

To investigate more carefully the differences between our AME approach
and the MC simulations on approaching the
transition, we analyse the local
arrangements of spins in the steady state. This is achieved by
equating the derivative of $\phi(t)$ in
Eq.~(\ref{eq:persistence_evolution}) to zero and exploring this and the master equations
to see the possible system configurations under which a non-zero value
of $\Phi$ is possible.  

It is evident that $\phi$ will be in the steady state
only if each of the $\phi_{m_1, m_2, m_3, m_4}^-$ and $\phi_{m_1,
  m_2, m_3, m_4}^+$ variables are also in the steady state. However, there
are no requirements for the $\psi_{m_1, m_2, m_3, m_4}^-$ and $\psi_{m_1,
  m_2, m_3, m_4}^+$ variables to be in a steady state, and indeed one of the
configurations of the system at equilibrium is a dynamical one
where the flipped nodes are still mobile and dynamically active. The
system configuration in this regime is summarized in Table~\ref{tab:nonzerovars}.

\begin{table}[t]
\begin{tabular}{| r | c |}
\hline
\multicolumn{2}{|c|}{Non-zero variables}  \\
\hline
$\bar{\phi}^-_{m_1, m_2, m_3, m_4}$ & $m_3=m_4=0$ and $m_1+m_3 < f$ \\
$\bar{\phi}^+_{m_1, m_2, m_3, m_4}$ & $m_3=m_4=0$ and $m_1+m_3 < f$ \\
$\bar{\psi}^-_{m_1, m_2, m_3, m_4}$ & $m_1=m_2=0$ \\
$\bar{\psi}^+_{m_1, m_2, m_3, m_4}$ & $m_1=m_2=0$ \\
\hline
\end{tabular}
\caption{The only values of $m_1, m_2, m_3$ and $m_4$ for which the different AME
  variables are non-zero in the dynamical steady state regime. Overbars denotes the steady state value.}
\label{tab:nonzerovars}
\end{table}

The other possible configuration of the system at equilibrium is one
where every node is immobile, being surrounded by less than $f$ spin-down
nodes. However, this configuration is highly unlikely for non-zero
values of $T$ and
furthermore it is not observed in the numerical simulations; we henceforth only
regard the dynamical steady state.

Analysis of the steady state equations for the dynamical equilibrium
yields the following conditions. The first is that  
\begin{equation}
\bar{\phi}_{m_1,m_2,m_3,m_4}^+=\bar{\phi}_{m_1,m_2,m_3,m_4}^-=0\;\;\;\;\; \forall \;
m_1+m_3\geq f.
\end{equation}
This simply states that the unflipped nodes can remain in the system but
only if they are
surrounded by less than $f$ spin-down nodes and so are
immobile. The second condition is on the neighbor
transition rate approximations. This condition is that all of these
rates are zero except for $\lambda_{3\rightarrow
  4}^{\psi^-},\lambda_{4\rightarrow 3}^{\psi^-},\lambda_{3\rightarrow
  4}^{\psi^+}$ and $\lambda_{4\rightarrow 3}^{\psi^-}$. These four
rates describe the transitions of flipped neighbours of
flipped nodes. The fact that they are non-zero in the steady state
regime of $\phi$, while the other transition rates are zero, indicates that the 4-state AME approach recreates
dynamical heterogeneity, a stylised fact of the glass transition~\cite{biroli2012random} where blocked
nodes and mobile nodes can co-exist when the system is in 
dynamical equilibrium. 

\begin{figure}[tb]
  \centering
  \vspace{5pt}
  \begin{subfigure}[b]{0.7\columnwidth}
    \includegraphics[width=0.9\textwidth]{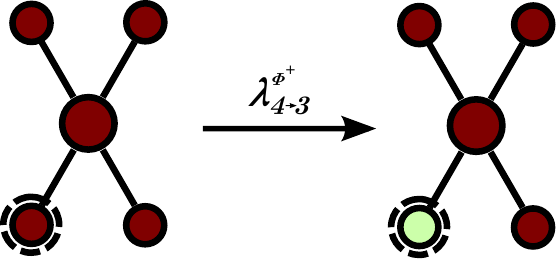}
    \caption{}
    \label{fig:interfacea}
  \end{subfigure}
  \begin{subfigure}[b]{0.7\columnwidth}
    \includegraphics[width=0.9\textwidth]{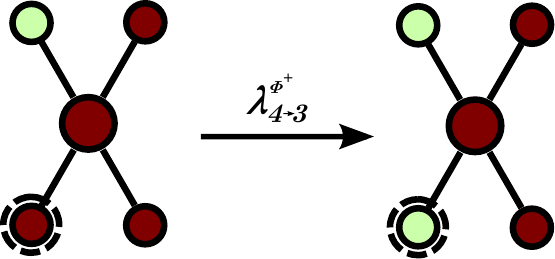}
    \caption{}
    \label{fig:interfaceb}
  \end{subfigure}
  \caption{(a) Example of a blocked spin at the interface which remains
    blocked regardless of its $\phi^+$ neighbor flipping. (b) Example
    of a blocked spin at the interface which becomes flippable after the
    change of its $\phi^+$ neighbor. Here, the facilitation parameter is
    $f=2$. In the AME, the two transition
    rates as shown by a) and b) here are the same as they are both
    approximated by $\lambda_{4\rightarrow 3}^{\phi^+}$.}
  \label{fig:interface}
\end{figure}

The neighbour transition rates are functions of the state variables as shown in
Eq.~(\ref{eq:4statetransitionrate}), and so for the transition rates
to satisfy the second condition it is required that
\begin{eqnarray}
\bar{\phi}_{m_1,m_2,m_3,m_4}^-=\bar{\phi}_{m_1,m_2,m_3,m_4}^+=0\;\;\;\;\; \forall \;
m_3, m_4 > 0, \\
\bar{\psi}_{m_1,m_2,m_3,m_4}^-=\bar{\psi}_{m_1,m_2,m_3,m_4}^+=0\;\;\;\;\; \forall \;
m_1, m_2 > 0.
\end{eqnarray}
This implies that there are no links between changed and unchanged
nodes. The reason that this
condition is necessary is because of the neighbor transition rate
approximation. This is illustrated in Fig.~\ref{fig:interface}, where
we show the two types of node-neighbor configurations that can appear at the
boundary and are observed in the MC simulations. Note in reality
that the flipped neighbor of the central node in
Fig.~\ref{fig:interfacea} will be able to flip without
releasing the cluster because the node has no other spin-down
neighbors. However, the flipped neighbor of the node in Fig.~\ref{fig:interfaceb} will not
be able to flip without releasing the cluster, as if it flips to
spin-down then the node will have sufficiently many spin-down
neighbors to flip. Therefore in reality, the neighbor transition rate $W(\phi_{0, 3, 0, 1}^+\rightarrow\phi_{0,3,1,0}^+)$ for the node in Fig.~\ref{fig:interfacea} should be
non-zero while the neighbor transition rate $W(\phi_{1, 2, 0, 1}^+\rightarrow\phi_{1,2,1,0}^+)$  in Fig.~\ref{fig:interfacea} should be zero. However,
the AME approximates neighbor transitions by link transitions, and in
this case the two transition rates are approximated by the same link
transition rate $\lambda^{\phi^+}_{4\rightarrow 3}$. This link transition rate is
necessarily zero to prevent the 
release of the nodes of type Fig.~\ref{fig:interfacea}. However, this link transition rate is
of the form of Eq.~(\ref{eq:4statetransitionrate}), and for its value to be zero it is
required that links of this type do not exist.

Thus the approximation of the neighbor
transition rates by the AME is compensated by the assumption that the size
of boundary between the blocked and mobile clusters is zero. It will be
now shown that it is this zero-boundary assumption that causes the inaccuracy of the AME in the $\alpha$-
relaxation regime as the size of the boundary, or in fact the size of the
critical clusters with large interface that compose it, diverges on
approaching the glass transition.

\section{Critical clusters}
\label{sec:criticalclusters}

Progress in approaching analytically the equilibrium properties of the FA model has been quite slow.
It took about 20 years, since the introduction of the model, for the steady states to be calculated on a locally tree-like network \cite{sellitto2005}.
Here we show that it is also possible to characterize the critical
clusters of the FA model by using a formalism recently developed in network percolation.
It has been noted \cite{toninelli2007jamming}, that the FA model is very similar to {\kcore} percolation and therefore the critical clusters of the FA model should correspond to the so-called corona clusters in {\kcore} percolation \cite{goltsev2006}.
However, the FA model is slightly more complex than {\kcore} percolation and,  to the best of our knowledge, it has not been shown explicitly that the critical clusters of the FA model can indeed be calculated following the same procedure used for {\kcore} percolation.
In this section we give a definition of critical clusters as a subset of the blocked clusters in the steady state and analytically prove that their mean size diverges at the phase transition.
As the found critical clusters are characterized by a large number of interface edges, this also explains why the quality of our AME approximation deteriorates at the phase transition.

In analogy with the corona clusters in {\kcore}  percolation \cite{goltsev2006}, we now consider as critical clusters the subsets of the blocked clusters where the minimum local requirement is exactly satisfied for all the nodes in the critical cluster.
In other words, a node belongs to such clusters if it is in a blocked cluster and it has exactly $(k-f+1)$ $\phi^+$-neighbors: the flipping of just one of the $\phi^+$-neighboring spins would create a cascade of movements that would eventually destroy the whole considered critical cluster at $t\to\infty$.
Our goal is to prove that these clusters are critical by showing that their mean size diverges at the transition.
In order to do that, we use the generating function formalism as in \cite{newman2003,goltsev2006}.
We define $H_{++}(x)$ as the generating function of the probability that following an edge in a blocked cluster from a $(\sigma=+1)$-spin, one gets to a spin $\sigma=+1$ node which belongs to a finite critical cluster.
Similarly, we define $H_{-+}(x)$ as the generating function of the probability that following an edge in a blocked cluster from a $(\sigma=-1)$-spin, one gets to a spin $\sigma=+1$ node which belongs to a finite critical cluster.
Then, the following equation holds
\begin{equation}
	H_{++}(x) = Q_{++} + x G_{++}(H_{++}(x)),
	\label{eq:H++}
\end{equation}
where
\begin{equation}
	Q_{++} = \rho \sum_{l=k-f+1}^{k-1} {k-1 \choose l} (Z_{++})^l (1-Z_{++})^{k-1-l},
	\label{eq:R++}
\end{equation}
\begin{equation}
	G_{++}(x) = \rho {k-1 \choose k-f}  (1-Z_{++})^{f-1} x^{k-f}.
	\label{eq:F++}
\end{equation}
Accordingly, $Q_{++}$ represents the probability that following an edge in a blocked cluster, one gets to a $\phi^+$-node which does not belong to a critical cluster because it has at least $(k-f+1)$ $\phi^+$-neighbors (so more than the minimum requirement), while $G_{++}(x)$ calculates the number of ways the $\phi^+$ edge endpoint can have exactly $(k-f)$ potential $\phi^+$-neighbors.
It is easy to notice that $G_{++}(Z_{++}) = Z_{++} - Q_{++}$ and $H_{++}(1)=Z_{++}$.
An analogous equation can be written for $H_{-+}(x)$.
The generating function $H_0(x)$ of the critical cluster sizes is, then, given by
\begin{equation}
	H_{0}(x) = \sum_{l=k-f+1}^{k} \left\{ P_+(l) \left[H_{++}(x)\right]^l + P_-(l) \left[H_{-+}(x)\right]^l \right\},
	\label{eq:H0}
\end{equation}
where
\begin{eqnarray}
	P_{+}(l) &=& \frac{ \rho {k \choose l} (Z_{++})^l (1-Z_{++})^{k-l} } {\Phi^{+}},\\
	P_{-}(l) &=& \frac{ (1-\rho) {k \choose l} (Z_{-+})^{l} (1-Z_{-+})^{k-l} } {\Phi^{-}}
\end{eqnarray}
are the degree distributions in blocked clusters for nodes with spins up and down, respectively.

The mean size of the critical clusters is given by $H'_{0}(1)$, which, being a linear combination of $H_{\pm+}(1)$ and $H'_{\pm+}(1)$, diverges at the phase transition only if the latter quantities do.
Therefore, as we have seen that $H_{++}(1)=Z_{++}$, we concentrate now on calculating $H'_{++}(1)$.
From (\ref{eq:H++}), we get
\begin{equation}
	H'_{++}(1) = \frac{ Z_{++} - Q_{++} }{ 1 - G'_{++}(Z_{++}) }.
	\label{eq:Hprime++}
\end{equation}
From the second condition of criticality (\ref{eq:Z++conditions}) and Eq.~(\ref{eq:F++}) we obtain
\begin{equation}
	\rho \sum_{l=k-f}^{k-2} {k-1 \choose l+1} (Z_{++})^l (1-Z_{++})^{k-l-2} = \frac{k-f-1}{k-f} G'(Z_{++}),
\end{equation}
from which
\begin{eqnarray}
	g(Z_{++}) &=& \frac{G'(Z_{++})}{k-f} + \rho \sum_{l=k-f}^{k-2} {k-1 \choose l+1} (Z_{++})^l (1-Z_{++})^{k-l-2} \nonumber\\
	&= & G'(Z_{++}).
\end{eqnarray}
Then, from the first condition of criticality (\ref{eq:Z++conditions}) we have $g(Z_{++})=G'_{++}(Z_{++})=1$.
Therefore, we have proved analytically that the investigated clusters are, indeed, critical, because at the phase transition their mean size diverges according to (\ref{eq:Hprime++}).

\section{Conclusions}
\label{sec:conclusions}

In this paper, we have introduced new analytical approaches to investigate both the steady state and the time relaxation of the Fredrickson-Andersen (FA) model.
Our analysis has then been compared with numerical simulations.
We have extended to a 4-state model an approximate master equation (AME) formalism~\cite{gleeson2013} to reproduce the dynamics of the model.
Unlike earlier theoretical approaches, our formalism is able to reproduce both the exact steady state and the transient regime.
In particular, we show that our approximation can partially capture dynamical heterogeneity, a characteristic of glassy systems where mobile and blocked clusters coexist. 
The degree of accuracy of the analytical approximation compared to the Monte Carlo is excellent in
general, save for a range of temperatures close to the
critical temperature.
We identify as a source of error the difficulty for the AME in capturing boundaries between blocked and flippable clusters.
To properly investigate this issue, we analytically identify the critical clusters of the model and show that at the glass transition the interface dominates the blocked clusters.
Therefore, also at $T \gtrsim T_c$ the dynamics should be largely affected by the slow unblocking of large quasi-critical clusters, with many interface edges that are not exactly captured by the AME.

There is much scope for progress in investigating this type of glass model using our
4-state AME approach. Here, the model was implemented on a degree-regular
network where each node had the same facilitation $f=2$.
Richer behavior occurs if the facilitation parameter value is allowed to vary
between nodes \cite{sellitto2010}.
In this framework, this is equivalent to considering a model with
uniform $f$ but where nodes do not all have the same
degree.
In other words, an appropriate definition of the degree distribution determines the model one may wish to study.

Degree variation in the 4-state AME formalism
can be naturally implemented through the degree distribution $p_k$.
Moreover, it is straightforward to extend the formalism we use to calculate the critical clusters in a network with a given degree distribution.

This work also paves the way for further analytical exploration of the model.
An expression for $\Phi$ can be obtained by taking the pair approximation
to the full system of master equations~\cite{gleeson2013}. Generating
functions can be used to reduce the system of master equation to a set
of ordinary differential equations~\cite{silk2013}.
Finally, mode coupling theory makes predictions
about the temporal evolution of glassy systems such as relationships
between relaxation time exponents. While the past study of this area was
restricted to examination of the MC simulations, our formalism
may give scope for analytical investigations.

This work has been partially funded by Science Foundation Ireland,
grant 11/PI/1026, and the FET-Proactive project PLEXMATH
(FP7-ICT-2011-8; grant 317614) funded by the European Commission. We acknowledge the DJEI/DES/SFI/HEA Irish Centre for High-End Computing (ICHEC) for the provision of computational facilities and support.

\appendix

\section{Binary State approach}
\label{sec:binmodel}

For the binary state approach, we only distinguish the spin state of
nodes. Therefore the model variables
are $\phi_{l,m}^-$ and $\phi_{l,m}^+$, the fraction of $-1$
(resp. $+1$) nodes in the network which have not previously flipped and which have $l$ neighbors in the state
$-1$ and $m$ neighbors in the state $+1$, for all values $l+m=k$ for
all possible $k$. In the same manner as described in
Section~\ref{sec:fourmodel}, master equations for $\phi_{l,m}^-$ and
$\phi_{l,m}^+$  can be constructed. The evolution equation for
$\phi_{l,m}^-$, before approximation of the neighbour transition rates
as in Eq.~(\ref{eq:link_approx}) in the 4-state case, is given by
\begin{multline}
\frac{d}{dt}\phi_{l, m}^- = -F(l)\phi_{l, m}^- \\
-W(\phi_{l,m}^-\rightarrow \phi_{l+1,m-1}^-)\phi_{l, m}^- -W(\phi_{l,m}^-\rightarrow \phi_{l-1,m+1}^-)\phi_{l, m}\\
+W(\phi_{l+1,m-1}^-\rightarrow \phi_{l,m}^-)\phi_{l+1, m-1}^- \\+W(\phi_{l-1,m+1}^-\rightarrow \phi_{l,m}^-)\phi_{l-1, m+1}.
\label{eq:binarymasterequation}
\end{multline}
with a similar equation for $\phi_{l,m}^+$. The evolution of the persistence
is then simply
\begin{eqnarray}
\frac{d}{dt}\phi &=&
\Big\langle\sum_{l=0}^k\frac{d}{dt}\phi_{l,m}^-+\frac{d}{dt}\phi_{l,m}^+\Big\rangle_k \nonumber
\\
=& -& \Big\langle\sum_{l=0}^k
F(l)\phi_{l,m}^-+R(l)\phi_{l,m}^+\Big\rangle_k.
\label{eq:persistence_evolution_binary}
\end{eqnarray}

Eq.~(\ref{eq:persistence_evolution_binary}), along
with the system of differential equations for $\phi_{l,m}^-$ and $\phi_{l,m}^+$ as given by
Eq.~(\ref{eq:binarymasterequation}), can be equated to zero to solve for
conditions yielding a non-zero value of $\Phi$ and thus the glassy state. The steady
state solution to Eq.~(\ref{eq:persistence_evolution_binary}) gives the
condition that $\phi_{l,m}^-$ is zero for $l \geq f$ and non-zero for
$l<f$, with the same condition for $\phi_{l,m}^-$. This is obvious, implying that
unflipped nodes can remain in the system but only if they are surrounded by
at most $f-1$ spin-down nodes.

Of more interest are the conditions on the neighbor transition
rates which arise from the steady state solutions to the differential equations for
$\phi_{l,m}^-$ and $\phi_{l,m}^+$. These conditions are that
\begin{eqnarray}
W(\phi_{l,m}^-\rightarrow \phi_{l+1,m-1}^-) &\mbox{\;\;\;\;}&
 \begin{cases} > 0 & 0 \leq l < f-1 \\
= 0 &   f-1 \leq l < k \end{cases}  \nonumber \\
W(\phi_{l,m}^-\rightarrow \phi_{l-1,m+1}^-) &\mbox{\;\;\;\;}&
 \begin{cases} > 0 & 0 < l \leq f-1 \\
= 0 &   f-1 < l \leq k \end{cases}
\label{eq:eqconditions}
\end{eqnarray}
with the same conditions for $W(\phi_{l,m}^+\rightarrow
\phi_{l+1,m-1}^+)$ and $W(\phi_{l,m}^+\rightarrow
\phi_{l-1,m+1}^+)$. These neighbor transition rate conditions
simply state that neighbors of blocked nodes can be mobile - however
the number of mobile neighbors is strictly less than $f$ and thus
sufficiently small that the blocked nodes will never become unblocked. Thus the model reproduces dynamical heterogeneity, a
stylised fact of the glass transition~\cite{biroli2012random} where blocked
nodes and mobile nodes can co-exist when the system is in
dynamical equilibrium.

As mentioned earlier, the neighbor
transition rates of the AME are not exact but rather approximated by mean field
link transition rates~\cite{gleeson2013}. For example, the second transition rate in
Eq.~(\ref{eq:binarymasterequation}) is approximated by
\begin{equation}
  W(\phi_{l,m}^-\rightarrow \phi_{l-1,m+1}^-)\approx l\beta^-
\label{eq:approx}
\end{equation}
where $\beta^-$ is the mean-field rate that a link of type
$(-1)$---$(-1)$ changes to $(-1)$---$(+1)$ and is given by
\begin{equation}
  \beta^- = \frac{\langle\sum_{l=0}^k l F(l)\phi_{l,m}^-\rangle_k}{\langle\sum_{l=0}^k l\phi_{l,m}^-\rangle_k};
\end{equation}
see~\cite{gleeson2013} for details. This is the level of approximation in the model. These mean-field rates fail to
capture the dynamic heterogeneities of the FA system. In particular, they
are always non-zero and so do not satisfy the neighbor transition
rate condition of Eq.~(\ref{eq:eqconditions}). This implies that a
non-zero value of $\Phi$ is impossible in the
binary-state AME for all values of the temperature
$T$ and so $\Phi \equiv 0$. This is not accurate, as the exact value of $\Phi$ is non-zero
for all $T<T_c$ as can be seen in Fig.~\ref{fig:steady-state}, and so
we conclude that a binary-state AME - accounting only for the spin of
each node - is not sufficent to capture the FA model.

\section{Full set of equations}
\label{sec:appendix2}

The full set of equations for the 4-state AME, as described in
Section~\ref{sec:fourmodel}, are given by

\begin{widetext}
  \begin{multline}
    \frac{d}{dt}\phi_{m_1, m_2, m_3, m_4}^- =
 -F(m_1+m_3)\phi_{m_1, m_2, m_3,
  m_4}^- \\
-m_1\lambda_{1\rightarrow 4}^{\phi^-}\phi_{m_1, m_2, m_3, m_4}^-
-m_2\lambda_{2\rightarrow 3}^{\phi^-}\phi_{m_1, m_2, m_3,
  m_4}^-
-m_3\lambda_{3\rightarrow 4}^{\phi^-}\phi_{m_1, m_2, m_3,
  m_4}^- -m_4\lambda_{4\rightarrow 3}^{\phi^-}\phi_{m_1, m_2, m_3,
  m_4}^- \\
+ (m_1+1)\lambda_{1 \rightarrow 4}^{\phi^-}\phi_{m_1+1, m_2, m_3,m_4-1}^-  +
(m_2+1)\lambda_{2 \rightarrow 3}^{\phi^-}\phi_{m_1, m_2+1, m_3-1,m_4}^- \\
+ (m_3+1)\lambda_{3 \rightarrow 4}^{\phi^-}\phi_{m_1, m_2, m_3+1,m_4-1}^-  +
(m_4+1)\lambda_{4 \rightarrow 3}^{\phi^-}\phi_{m_1, m_2,
  m_3-1,m_4+1}^-
\label{eq:masterequation1}
\end{multline}
\begin{multline}
    \frac{d}{dt}\phi_{m_1, m_2, m_3, m_4}^+ =
 -R(m_1+m_3)\phi_{m_1, m_2, m_3,
  m_4}^+ \\
-m_1\lambda_{1\rightarrow 4}^{\phi^+}\phi_{m_1, m_2, m_3, m_4}^+
-m_2\lambda_{2\rightarrow 3}^{\phi^+}\phi_{m_1, m_2, m_3,
  m_4}^+
-m_3\lambda_{3\rightarrow 4}^{\phi^+}\phi_{m_1, m_2, m_3,
  m_4}^+ -m_4\lambda_{4\rightarrow 3}^{\phi^+}\phi_{m_1, m_2, m_3,
  m_4}^+ \\
+ (m_1+1)\lambda_{1 \rightarrow 4}^{\phi^+}\phi_{m_1+1, m_2, m_3,m_4-1}^+  +
(m_2+1)\lambda_{2 \rightarrow 3}^{\phi^+}\phi_{m_1, m_2+1, m_3-1,m_4}^+ \\
+ (m_3+1)\lambda_{3 \rightarrow 4}^{\phi^+}\phi_{m_1, m_2, m_3+1,m_4-1}^+  +
(m_4+1)\lambda_{4 \rightarrow 3}^{\phi^+}\phi_{m_1, m_2,
  m_3-1,m_4+1}^+
\label{eq:masterequation2}
\end{multline}
\begin{multline}
    \frac{d}{dt}\psi_{m_1, m_2, m_3, m_4}^- =
 -F(m_1+m_3)\psi_{m_1, m_2, m_3,
  m_4}^-+R(m_1+m_3)\phi_{m_1, m_2, m_3,
  m_4}^++R(m_1+m_3)\psi_{m_1, m_2, m_3,
  m_4}^+ \\
-m_1\lambda_{1\rightarrow 4}^{\psi^-}\psi_{m_1, m_2, m_3, m_4}^-
-m_2\lambda_{2\rightarrow 3}^{\psi^-}\psi_{m_1, m_2, m_3,
  m_4}^-
-m_3\lambda_{3\rightarrow 4}^{\psi^-}\psi_{m_1, m_2, m_3,
  m_4}^- -m_4\lambda_{4\rightarrow 3}^{\psi^-}\psi_{m_1, m_2, m_3,
  m_4}^- \\
+ (m_1+1)\lambda_{1 \rightarrow 4}^{\psi^-}\psi_{m_1+1, m_2, m_3,m_4-1}^-  +
(m_2+1)\lambda_{2 \rightarrow 3}^{\psi^-}\psi_{m_1, m_2+1, m_3-1,m_4}^- \\
+ (m_3+1)\lambda_{3 \rightarrow 4}^{\psi^-}\psi_{m_1, m_2, m_3+1,m_4-1}^-  +
(m_4+1)\lambda_{4 \rightarrow 3}^{\psi^-}\psi_{m_1, m_2,
  m_3-1,m_4+1}^-
\label{eq:masterequation3}
\end{multline}
\begin{multline}
    \frac{d}{dt}\psi_{m_1, m_2, m_3, m_4}^+ =
 -R(m_1+m_3)\psi_{m_1, m_2, m_3,
  m_4}^++F(m_1+m_3)\phi_{m_1, m_2, m_3,
  m_4}^-+F(m_1+m_3)\psi_{m_1, m_2, m_3,
  m_4}^- \\
-m_1\lambda_{1\rightarrow 4}^{\psi^+}\psi_{m_1, m_2, m_3, m_4}^+
-m_2\lambda_{2\rightarrow 3}^{\psi^+}\psi_{m_1, m_2, m_3,
  m_4}^+
-m_3\lambda_{3\rightarrow 4}^{\psi^+}\psi_{m_1, m_2, m_3,
  m_4}^+ -m_4\lambda_{4\rightarrow 3}^{\psi^+}\psi_{m_1, m_2, m_3,
  m_4}^+ \\
+ (m_1+1)\lambda_{1 \rightarrow 4}^{\psi^+}\psi_{m_1+1, m_2, m_3,m_4-1}^+  +
(m_2+1)\lambda_{2 \rightarrow 3}^{\psi^+}\psi_{m_1, m_2+1, m_3-1,m_4}^+ \\
+ (m_3+1)\lambda_{3 \rightarrow 4}^{\psi^+}\psi_{m_1, m_2, m_3+1,m_4-1}^+  +
(m_4+1)\lambda_{4 \rightarrow 3}^{\psi^+}\psi_{m_1, m_2,
  m_3-1,m_4+1}^+
\label{eq:masterequation4}
\end{multline}\end{widetext}
with initial conditions
\begin{align}
\psi_{m_1,m_2,m_3,m_4}^-(0)&=0 \\
\psi_{m_1,m_2,m_3,m_4}^+(0)&=0 \\
\phi_{m_1,m_2,m_3,m_4}^-(0)&= \begin{cases}
  p_k(1-\rho)\binom{k}{m_1}(1-\rho)^{m_1}\rho^{m_2} &\mbox{if }
  m_3=m_4=0 \\ 
0 & \mbox{otherwise} \end{cases} \\
\phi_{m_1,m_2,m_3,m_4}^+(0)&= \begin{cases}
  p_k~\rho\binom{k}{m_1}(1-\rho)^{m_1}\rho^{m_2} &\mbox{if }
  m_3=m_4=0 \\ 
0 & \mbox{otherwise} \end{cases}
\end{align}
and where $F$ and $R$ are defined as
\begin{align}
  F(m_1+m_3) &=
  \begin{cases} 0 &\mbox{if } m_1+m_3 < f \\
1 & \mbox{if } m_1+m_3 \geq f \end{cases} \label{eq:ratefunctionF2}
\\
  R(m_1+m_3) &=
  \begin{cases} 0 &\mbox{if } m_1+m_3 < f \\
e^{-1/T} & \mbox{if } m_1+m_3 \geq f. \end{cases}
\label{eq:ratefunctionR2}
\end{align}

\bibliography{library}

\end{document}